%% file: ms.tex
\documentclass[12pt,preprint]{aastex}

\slugcomment{}

\shorttitle{AO Observations of Protostars}
\shortauthors{Connelley et al.}

\begin{document}

\title{An Adaptive Optics Survey For Close Protostellar Binaries\altaffilmark{1}}

\author{Michael S. Connelley\altaffilmark{2}, Bo Reipurth\altaffilmark{3}, and Alan T. Tokunaga\altaffilmark{4}}

\altaffiltext{1}{Some of the data presented herein were obtained at the W.M. Keck Observatory, which is operated as a scientific partnership among the California Institute of Technology, the University of California and the National Aeronautics and Space Administration. The Observatory was made possible by the generous financial support of the W.M. Keck Foundation. Based in part on data collected at Subaru Telescope, which is operated by the National Astronomical Observatory of Japan.}
\altaffiltext{2}{NASA Ames Research Center, MS 245-6, Moffett Field, CA 94035}
\altaffiltext{3}{University of Hawai'i Institute for Astronomy, 640 N. Aohoku Pl., Hilo HI 96720}
\altaffiltext{4}{University of Hawai'i Institute for Astronomy, 2680 Woodlawn Dr., Honolulu, HI 96822}

\begin{abstract}

   In order to test the hypothesis that Class I protostellar binary stars are a product of ejections during the dynamical decay of non-hierarchical multiple systems, we combined the results of new adaptive optics (AO) observations of Class I protostars with our previously published AO data to investigate whether Class I protostars with a widely separated companion (r$>$200~AU) are more likely to also have a close companion (r$<$200~AU).  In total, we observed 47 embedded young stellar objects (YSOs) with either the Subaru natural guide star AO system or the Keck laser guide star AO system.  We found that targets with a widely separated companion within 5,000~AU are not more likely to have a close companion.  However, targets with another YSO within a projected separation of 25,000~AU are much more likely to have a close companion.  Most importantly, \emph{every} target with a close companion has another YSO within a projected separation of 25,000~AU.  We came to the same conclusions after considering a restricted sample of targets within 500~pc and close companions wider than 50~AU to minimize incompleteness effects.  The Orion star forming region was found to have an excess of both close binaries and YSOs within 25,000~AU compared to other star forming regions.  We interpret these observations as strong evidence that many \emph{close} Class I binary stars form via ejections and that many of the ejected stars become unbound during the Class I phase.  

\end{abstract}


%% subject header.  Objects should be in the appropriate "individual"
%% headers (e.g. quasars: individual, stars: individual, etc.) with the
%% additional provision that the total number of headers, including each
%% individual object, not exceed six.  The \objectname{} macro, and its
%% alias \object{}, is used to mark each object.  The macro takes the object
%% name as its primary argument.  This name will appear in the paper
%% and serve as the link's anchor in the electronic edition if the name
%% is recognized by the data centers.  The macro also takes an optional
%% argument in parentheses in cases where the data center identification
%% differs from what is to be printed in the paper.

\keywords{instrumentation: adaptive optics, stars: formation, binaries: visual}


\section{Introduction}

   It is well known that the majority of main-sequence stars are in binary systems.  For solar-type stars, the binary fraction over all separations is 61\% after correcting for incompleteness \citep{Duq1991}.  Observations of young binary stars are therefore critical to understanding the problem of star formation.  To this end, there have been numerous studies of T Tauri binary properties, summarized by \citet{Mat2000}.  Overall, the T Tauri binary fraction is roughly twice the solar-type main-sequence binary fraction \citep{Ghe2001}.  T Tauri stars also have more binary stars at wider separations than solar-type main-sequence stars \citep{Koh1998}.  The optical survey of 238 T Tauri stars by \citet{Rei1993} found 37 binaries with separations from 1\arcsec~ to 12\arcsec, but only 1 triple.  They propose that many of these binaries have a companion too close to resolve, and that these may be resolved with higher angular resolution observations.

  Surveys of the Class I binary fraction have only recently been conducted.  Observations by \citet{Con2008a}, \citet{Duc2004}, and \citet{Hai2004} show that Class I YSOs also have a binary fraction excess over solar-type main-sequence stars and have a similar binary fraction and binary separation distribution beyond $\sim$100~AU as T Tauri stars.  Most recently, \citet{Duc2007} reported on a search for binary companions among 45 embedded young stars with AO, and found a binary frequency of 47\%$\pm$8\% over the separation range of 14~AU to 1400~AU.  \citet{Rei2000} presented a scenario by which most stars are born in higher order multiple systems that dynamically decay.  In order to explain the evolution of the Class I binary separation distribution, \citet{Con2008b} adopted this hypothesis and proposed that the envelope that remains after the Class 0 phase helps to keep ejected companions from becoming unbound until the envelope dissipates.  These scenarios predict that Class I YSOs with widely separated binary companions should also have a much closer companion.

   Adaptive optics is a maturing technology that has been shown to be particularly useful for investigating the binary properties of young stars.  \citet{Bec2003} observed the NGC~2024 embedded cluster with AO at K-band.  Of 73 stars observed, they found only 3 binaries and 1 triple, which is consistent with the solar-type main-sequence binary fraction within the separation range (145~AU to 950~AU) in which they were sensitive to companions.  Twenty-four nearby pre-main-sequence stars were observed with AO by \citet{Bra2003}, who found that 10 of these are close binaries and 2 are triple star systems.  More recently, \citet{Cor2006} used AO to observe known binary T Tauri stars, searching for higher order multiples.  Over the separation range of 12~AU to 2000~AU, they found that 27\% of the total number of systems they observed are higher order multiples, including 7 triple and 7 quadruple star systems.  Laser guide star AO is a new technology that allows AO observations of many more embedded sources than was previously possible with natural guide star AO.

   The goal of this study is to explore the possibility that many widely separated Class I binary stars are the result of an ejection. \citet{Con2008b} predicted that protostars with widely separated companions within a projected separation of 5,000~AU should be more likely to have a close companion than protostars without a widely separated companion.  If so, then Class I objects with a closely separated companion may also be more likely to have a widely separated companion than Class I objects without a closely separated companion.  We also consider nearby YSOs that may have been ejected but are now too widely separated to be a gravitationally bound companion.  In this paper we describe our sample selection and observations (Section 2), analyze our results (Section 3) and discuss our findings and what we have learned about star formation (Section 4).

\section{Sample Selection \& Observations}

   This paper uses the AO observations presented in \citet{Con2008a} taken with the Subaru Telescope as well as new observations taken with the Keck II telescope.  The sample of Class I YSOs that we observed with AO was selected from the protostars listed in Table 1 of \citet{Con2008a}.  The Subaru natural guide star adaptive optics (NGS-AO) system requires that a R$<$16 guide star be within $\sim$30\arcsec~ of the target whereas the Keck II laser guide star adaptive optics (LGS-AO) system requires that a R$<$18 tip-tilt reference star be within $\sim$40\arcsec~ of the target.  We used the USNO B1.0 catalog \citep{Mon2003} to find visible guide stars within these radii from the list of Class I YSOs for the appropriate observing run.  We also visually inspected POSS plates to verify that the catalog entry is not a false detection or a reflection nebula.   In the case of our Keck II observations, we also used the Keck Observatory's AO planning tool to ensure that the guide star was in a location where the wavefront sensor could use it.  In addition to selecting targets based on the availability of natural guide stars, we chose sources that are redder than H$-$K$=1$ to ensure that the source is deeply embedded.  We also avoided sources with significant reflection nebulosity at K-band since structure in the nebula could interfere with detecting a companion.  

   We used the 36 element curvature NGS-AO system and CIAO (Murakawa et al. 2004) with the 0\farcs0213 pixel$^{-1}$ plate scale on the Subaru telescope on the nights of Nov. 15 and 16, 2005.  All targets were observed at L$'$ (3.8~$\mu$m). The median FWHM angular resolution of our Subaru AO L$'$ data is 0\farcs138 with a standard deviation of 0\farcs035 (the diffraction limit is 0\farcs096).  This observing run focused on targets in the Taurus and Orion star forming regions.  We were not able to observe all of the targets we intended to observe since a full moon in Taurus prevented us from observing many targets in that region and our third night was too cloudy to use the AO system.  We used the Keck II LGS-AO system and NIRC2 with the 0\farcs01 pixel$^{-1}$ plate scale camera during two half-nights, on June 10 and July 9, 2007.  Our observations were conducted primarily at L$'$.  The median FWHM angular resolution of our Keck L$'$ data is 0\farcs087 with a standard deviation of 0\farcs014, which compares favorably to the diffraction limit of 0\farcs078.  Many of the close binaries found with our L$'$ observations were also observed at K-band. 

     Dithering was used for all observations in order to remove the detector flat field effects.  In the case of L$'$ observations, coadds were used to increase the effective integration time per image to $\sim$20~s to increase observing efficiency.  For the Subaru observations, three images were taken at each dither position to reduce the effect of the overhead involved with dithering.  For the Keck observations, we stayed at each dither position long enough for the low bandwidth wavefront sensor to take a few exposures (which varied with the brightness of the tip-tilt reference star).  Standard stars that have been observed by UKIRT through the MKO filter set (Simons \& Tokunaga 2002, Tokunaga \& Simons 2002) were selected from the UKIRT faint standard star list, and were observed for photometric calibration.  

   All data were reduced using the same procedure.  A dark frame was made by averaging together 10 individual dark frames of the same exposure time as the science data.  This dark was then subtracted from each target frame.  To make the ``sky'' frame, each dark subtracted frame was scaled to have the same median value, then averaged together using a min-max rejection.  The resulting sky frame was then normalized using the median value of the pixel counts.  Each dark subtracted (but not scaled) target frame was divided by this normalized ``sky flat''.  The median sky value for each frame was subtracted from each frame to set the average background counts in each frame to 0.  The images were then aligned and averaged together using an average sigma clipping rejection.  Since the AO optics are warm, nearly all of the L$'$ ``sky'' brightness is from the telescope and the AO system.  The procedure we used did not make a true L$'$ flat.  However, the L$'$ ``flats'' we used were effective in removing the flat-field properties due to the detector, which dominates over the signal of the target in the raw data.  Figure 1 shows the 3\farcs0$\times$3\farcs0 area around each binary star observed with Keck.  AO images taken with Subaru are presented in \citet{Con2008a}.

   We subtracted the PSF of the primary star to look for close and faint companions.  Nearby field stars made the best PSF reference stars.  However, given the narrow field of view and the scarcity of bright stars at L$'$, these were rare.  In the case of the Subaru observations, a radial average of the primary star's PSF was also used as an effective PSF model.  This technique could not be used for the Keck observations due to the hexagonal pupil, so we often used stars that were observed just before or after the target.  However, constructing a PSF from other target stars was difficult due to how the AO correction depended on the brightness of the tip-tilt reference star and the rotation of the pupil from target to target.  The Keck PSF was also not point symmetric, so rotating the PSF 180$^{\circ}$ then subtracting resulted in large residuals.
  
\subsection{Separation and Position Angle Uncertainties}
   Since both Keck II and Subaru are altitude-azimuth mounted telescopes, the instrument rotator (in the case of Subaru) or field of view rotator (for Keck II) keeps the orientation of the field of view fixed and also sets the position angle of the field of view.  We have noted discrepancies of 15$^{\circ}$ and 10$^{\circ}$, respectively, in the position angles of binaries between our Keck and Subaru data of IRAS 21004+7811 and between our Keck and UKIRT data of IRAS 18383+0059.  Since UKIRT is equatorially mounted, the orientation of the field of view is fixed, so the PA measured from the UKIRT image is believed to be accurate.  Since these were the only cases among our Keck data where the field of view has North left and East down, it is likely that we neglected to set the rotator PA before we started the observation.  All PA measurements from our Keck and Subaru data are only as accurate as the image or instrument rotator, which are 10\arcsec ~for Subaru and conservatively estimated as $\sim$36\arcsec ~for Keck (private communications).  The uncertainties on both separation and PA are based on a 1/4 pixel centroiding accuracy.

\subsection{Results}

   Our observations are summarized in Table 1, including details of the binary companions observed in the course of our AO observations.  The observations in November 2005 were with the Subaru telescope, whereas the June and July 2007 observations were with the Keck II telescope.  The new binary companions discovered with the Keck II observations are IRAS 21004+7811 Ba and Bb, IRAS 21007+4951 B, and IRAS 21445+5712 B (see note in Appendix A).  Most of the other binary companions in Table 1 were found by \citet{Con2008a}.  FS Tau Aa and Ab were discovered by \citet{Sim1992}, and the 1\farcs82 companion to IRAS 06382+1071 was discovered by \citet{Pic1995}.

\section{Analysis}
\subsection{Questions to Test the Hypothesis}
   To address the goals of this study as stated in Section 1, we asked these questions: 1)  Are Class I protostars with a widely separated companion more likely to have a close companion (r$<$200~AU) than a protostar without a widely separated companion?  2) Are Class I protostars with a close companion more likely to have a wide companion? 3) Are Class I protostars without a close companion less likely to have an apparently widely separated companion?  4) Is there a regional dependence on the fraction of close or wide binary companions?  To answer these questions, we used an outer separation limit of 5,000 AU (within which a companion might be gravitationally bound) and 25,000~AU (at such a separation an object is unlikely to be gravitationally bound but is close enough that it may have been ejected, as described in Section 3.3).  The answers to these questions and their implications are discussed in section 4.

\subsection{Biases and Target Identification}
%% Paper includes only targets observed w/ AO
   The analysis in this paper only includes targets observed with AO.  Of the targets we observed with AO, 5 are in Taurus, Perseus, Auriga, and Ophiuchus, 14 are in Orion, and 29 are elsewhere.  In comparison, the study by \citet{Con2008a} includes 40 targets in Taurus, Perseus, Auriga, and Ophiuchus, 33 in Orion, and 152 are elsewhere.  The results of this study are therefore biased in favor of targets in the Orion star forming region and against targets in the Taurus, Perseus, Auriga, and Ophiuchus clouds.

   We compared the spectral index distributions (Figure 2) of the targets in this study with those in \citet{Con2008a} to determine if the AO observed sample of Class I YSOs is significantly different than their sample of Class I YSOs.  The median spectral index for all targets observed by \citet{Con2008a} is +0.79, with a standard deviation of 0.74, whereas for this study the median spectral index is +0.50 with a standard deviation of 0.48.  While this study is thus slightly biased towards more evolved YSOs, the difference is not statistically significant.  The 2-Sample Kolmogorov-Smirnov test shows that the two spectral index distributions are consistent with each other within 3~$\sigma$ confidence limits.  The AO observed sample is likely to be biased towards more evolved sources since many of the NGS-AO guide stars were cases where enough visible light from the embedded YSOs themselves was available to guide on.  Additional observations of more highly embedded sources are needed to address this bias.  

%% Asked Bo if optically visible stars can be included since they may have been ejected.
   Not all of the targets observed for this study are Class I YSOs (i.e. have an envelope, are not optically visible, etc.), as seen by the fact that some of the targets in this study were used as the optical guide star for the AO system.  Since IRAS 22376+7455 ($\#$4 and $\#$5)\footnote{The numbers refer to the identifications of the sources in Figure 6 of Connelley et al. (2008a)}, IRAS 05384-0807 ($\#$6), and IRAS 19247+2238 $\#3$ are optically visible stars and are not deeply embedded, and thus are not included in this analysis.  However, based on their proximity to the embedded protostars and optically visible reflection nebulosity, they appear to be associated with the dark cloud hosting the Class I objects.  These optically visible stars may be as young as the nearby embedded protostars and may have become optically visible after being ejected.  \citet{Rei2000} proposed that ejections can rapidly change the classification of a YSO, and the large number of hierarchical binaries in these three systems (9 in all) suggests that ejections played an important role in the dynamical histories of these IRAS sources.

\subsection{Binary Separation Ranges and Contamination}
%% Want to include "nearby" YSOs
     For the purpose of this analysis, a companion with a projected separation less than 200~AU is considered to be ``close'', and if greater than 200~AU is considered to be widely separated.  The key result of this study, that every embedded protostar with a close companion also has another YSO within a projected separation of 25,000~AU (see below), would not be changed if the definition of a ``close'' companion was as great as 500~AU. 

  Ejected stars may be gravitationally bound or unbound \footnote{An ejection is the dynamical process that changes a non-hierarchical system into a bound or unbound hierarchical system.  An ejection may or may not lead to the escape of the ejected member from the system.  A bound hierarchical system is not always stable and may lead to a later ejection and possible escape.}, and the decline in the number of binary companions with separations greater than 1000~AU versus spectral index suggests that many widely separated companions become unbound during the Class I phase \citep{Con2008b}.  To demonstrate that this scenario is plausible, consider an object ejected at the escape velocity of a 2~M$_{\odot}$ central (binary) star.  The ejected object would reach a distance of 25,000~AU in about $2\times10^{5}$~years, which is consistent with the expected Class I life time.  Objects with a projected separation beyond 5,000~AU are most likely unbound (whether or not they were ejected).  If these nearby unbound objects are at such distances on account of having been ejected, then these objects were likely to have been ejected at a velocity higher than the escape velocity, and thus can get to a separation of 25,000~AU in less than $2\times10^{5}$~years.  If our hypothesis is correct, then we can expect to find ejected YSOs (Class I and otherwise) at separations as great as 25,000~AU from our Class I targets.  YSOs at separations beyond 5,000~AU are considered as being ``nearby'' but are not considered to be a gravitationally bound binary companion.

  The distance to each target (when known) is given in Table 1.  We used the distance estimate to the Orion star forming region by \citet{San2007}, who measured the parallax of a star in the Orion Nebula Cluster to derive a distance estimate of $389^{+24}_{-21}$~pc.  To date, the most commonly used distance to the Orion star forming region was $\sim$470~pc (see discussion in Muench et al. 2008).  Using either distance estimate has no affect on the outcome of our analysis.

%% Comments on contamination
   Contamination must always be carefully considered when looking for binary companions.  In the case of our close binaries, the chance of contamination is negligibly small within a projected separation of 200~AU (0.02\% for targets in Taurus, Perseus, Auriga, and Orion).  This value is based on the density of background stars within $\Delta$L$'$=4 magnitudes of the primary stars in these parts of the sky.  For our more widely separated companion stars, contamination is a much greater problem since we are also considering YSOs within a projected radius of 25,000~AU.  We were often able to use the H, K, and L$'$ photometry in \citet{Con2008a} to determine if a candidate YSO within a projected separation of 25,000~AU has colors consistent with a YSO or a reddened background star.  In some cases we also used the presence of an optical or near-IR reflection nebula \citep{Con2007a} to ensure that the candidate YSO is in the same cloud.

    It is important to note that this study is different from many previous studies of binary stars that attempt to determine how many companions each star has.  We were not trying to determine how many companions a star has out to a separation of 25,000~AU, in which case we would have had to determine whether each object within that projected separation is a YSO or a background star.  Rather, we were only looking to see if \emph{any} object within this projected separation has properties (reflection nebula or near-IR colors) consistent with a YSO.  Thus, contamination by background stars is much less of a problem for this study, allowing us to extend the outer search radius to 25,000~AU.

\subsection{Restricted Sample}
%% No incompleteness correction, restricted sample
   Surveys for binary stars often include a correction to compensate for an incomplete sensitivity to close and faint binary companions.  This is particularly important when data were taken under different seeing conditions or targets are at different distances, as is the case here.  We did not use an incompleteness correction for this analysis since the present study is not investigating the number of binaries in a given separation range, but rather we are interested in whether individual targets have a binary companion and, if so, at what separation.  Thus, in addition to considering all companions to all targets, we considered a \emph{restricted sample} consisting of only targets within 500~pc and discarding binary companions with a projected separation of less than 50~AU to minimize incompleteness the effect of our incomplete sensitivity to close companions of more distant targets.  We chose these values since we are typically sensitive to all companions as closely separated as 0\farcs1 (50~AU at 500~pc) with a contrast less than $\Delta$L=3 with our AO data.  However, we searched for companions as faint as $\Delta$L=4, where we could typically detect companions only as close as 0\farcs34 (50~AU in Taurus or Ophiuchus, 160~AU at Orion).  Thus, we are typically not sensitive to any ``close'' companions if they are as distant as Orion and are more than $\Delta$L=3 magnitudes fainter than their primary star, but we could detect them in the closer star forming regions.  There are also two targets in the restricted sample (IRAS 22376+7455 $\#$2 and $\#$6) where our L$'$ observations were not deep enough to detect a companion 4 magnitudes fainter than the primary star.  

\subsection{Triple Systems}
%% About triples, if the companion is around the primary or secondary
   Among the 6 triple systems that we observed with AO, 4 have a companion to the primary star and 2 have a companion to the secondary\footnote{The primary star is defined as the star that is brightest at L$'$}.  The two cases of a binary secondary are IRAS~18383+0059 and IRAS~21004+7811.  In the case of IRAS~18383+0059, the secondary is just slightly fainter than the primary at L$'$ and is redder, so it is likely to be more luminous and thus may actually be the primary star in the system.  In the case of IRAS~21004+7811, all three stars have similar colors, and the primary is much brighter than the companions.  Our finding that 4 (or 5) of 6 triples have a binary primary star is consistent with the expectation that the least massive star is the most likely to be ejected when a non-hierarchical triple system decays via an ejection.  

\section{Discussion}

   The following discussion of the answers to our four questions uses results based on the restricted sample unless otherwise noted.  The data supporting this discussion, from both the restricted and unrestricted samples, are presented in Table 2.

   \emph{Question $\#$1: Are Class I protostars with a widely separated companion more likely to have a close companion than a protostar without a widely separated companion?}  Our hypothesis predicts that protostars with a widely separated companion should be more likely to have a close companion. We find that a nearly equal fraction of Class I protostars, $\sim30\%$, have a close companion whether or not they have a widely separated companion within 5000~AU.  This suggests that the answer to this question is no, contrary to our hypothesis.  However, when the outer separation limit is increased to 25,000~AU, we found that among the protostars in either the whole sample or the restricted sample that do not have a widely separated YSO, \emph{none} have a close companion.  In comparison, $\sim47\%$ of those protostars in the restricted sample with such a widely separated companion also have a close companion. Using the binomial confidence estimator in Appendix B.2 of \cite{Bra2006}, this is a higher percentage with a 99.6\% (2.63~$\sigma$) confidence. The results based on the unrestricted sample are consistent with the results from the restricted sample.

   \emph{Question $\#$2: Are Class I protostars with a close companion more likely to have a widely separated companion?}  If our hypothesis is correct, then Class I protostars with a close companion should be more likely to have a widely separated companion.  Considering companions out to a separation of 5000~AU, we find that this is true, but not by a statistically significant margin.  Only a third of those Class I protostars with a close companion also have a companion within 5000~AU.  However, we found that \emph{every} Class I protostar in either sample that has a close companion also has another YSO within 25,000~AU.  When widely separated YSOs are considered as far as 25,000~AU, Class I protostars with a close companion are more likely to have a widely separated YSO nearby with a confidence of $\sim3.9~\sigma$.  Recall that companions within 5,000 AU are considered bound, and companions within 25,000 AU to be unbound.  If our hypothesis is correct, then this result suggests that most (about two-thirds) of ejected companions become unbound by the end of the Class I phase.

   \emph{Question $\#$3: Are Class I protostars apparently without a close companion less likely to have an apparently widely separated YSO?}  If widely separated companions are created via ejections, then those Class I protostars without a close companion should tend to not have a widely separated companion.  Considering widely separated companions as far as 5,000~AU, we find that this is indeed the case in both samples.  However, in both samples, we found that Class I protostars without a close companion are equally likely to have or not to have another YSO within 25,000~AU.  In the cases where a Class I protostar does not have a resolved close companion but does have another YSO within 25,000~AU, it is possible that the Class I protostar does have a very close companion that remains unresolved, or the other YSO is within a projected separation of 25,000~AU by chance.
   
   \emph{Question $\#$4: Is there a regional dependence on the fraction of close or wide binary companions?}  \citet{Con2008b} considered the Class I binary separation distribution for targets in three groups.  Group I includes targets in the Orion star forming region. Group II includes targets in the Taurus, Auriga, Perseus, and Ophiuchus clouds.  Group III includes all targets in other clouds.  They found that the Class I binary separation distributions are significantly different between Group I and II.  Since we observed only a few targets in Group II with AO, and since the sample observed in this study is much smaller than the sample used by \citet{Con2008b}, we divided the AO observed sample between those targets in the Orion star forming region and those that are not in Orion.  We stress that the targets in Orion are not in the Orion Nebula Cluster, but are distributed throughout the Orion star forming region.  

   We found that the Orion and non-Orion sub-samples have the same fraction of widely separated companions within 5000~AU.  However, the Orion sub-sample has a significant excess of YSOs within 25,000~AU of the Class I target.  We find that there is a 92.5\% chance (1.4~$\sigma$) that the Orion and non-Orion sub-samples are different based on these data.  We found that the Orion sub-sample also has a significant excess of close binary stars versus the non-Orion sub-sample ($50.0\%^{+17.6\%}_{-17.6\%}$ and $15.0\%^{+12.4\%}_{-8.0\%}$, respectively).  The Orion and non-Orion sub-samples have a 95.1\% chance (1.7~$\sigma$) of being different based on this data. The results from the unrestricted sample are consistent with these findings.  The fact that the Orion sub-group has these excesses of close binaries \emph{and} nearby YSOs is consistent with the finding that the Orion star forming region (Group I) has a higher Class I binary frequency than Groups II or III ($17.6\%\pm8.2\%$ vs. $6.1\%\pm3.4\%$ vs. $8.2\%\pm3.4\%$ respectively, Connelley et al. 2008b).  This finding is also consistent with the hypothesis that young binary stars form via ejections from non-hierarchical multiple groups and that the nearby YSOs were ejected during the event that created the close binary.  It is unclear what cloud properties cause the Orion sub-sample to have a higher binary frequency.  
   
%% implications, what it all means
   These results are consistent with the scenario presented by \citet{Rei2000} in which a large fraction of Class 0 protostars form in non-hierarchical multiple systems that dynamically decay during the Class 0 phase.  However, it is important to note that not all widely separated binaries are necessarily the result of an ejection.  It is possible for a companion to have formed at a wide separation, or the presumed companion may be a YSO from elsewhere, or it may be a background star along the line of sight.  Our interpretation of the observations is that each close binary that we found is the remnant of a multiple system that ejected a member, and that the YSOs that are found within 25,000~AU are likely to have been ejected.  We also found that, among protostars with a close companion, a third have a companion within 5,000~AU (which is likely to be gravitationally bound) whereas all have a YSO within 25,000~AU (which is not likely to be bound if it is beyond 5,000~AU).  This result may suggest that about a third of the ejected stars are still bound in the Class I phase and the other two-thirds have become unbound.   

   \citet{Con2008b} found that the Class I binary fraction at separations between 1000~AU and 5000~AU steadily declines throughout the Class I phase (using spectral index as a proxy for age), presumably because these widely separated companions become unbound as the YSOs age. However, this result is for all Class I YSOs, not just those with a close resolved binary companion.  Under this scenario, we can expect that Class I YSOs with a close companion and a widely separated companion within 5,000~AU will be younger (and thus have a higher spectral index) than those Class I YSOs with a close companion but without a widely separated companion within 5000~AU (which has presumably become unbound) since the dissipation of the envelope allows widely separated companions to become unbound as the stars evolve.  The spectral indices (measured from 12~$\mu$m to 100~$\mu$m with IRAS fluxes) for those Class I YSOs with both a close companion and a widely separated companion within 5,000~AU have a median value of $0.62\pm0.38$, whereas the spectral indices for those with both a close companion but without a wide companion within 5,000~AU have a median value of $1.03\pm0.43$.  The former group only includes 6 IRAS sources, and the latter group 4 IRAS sources.  This result, strongly limited by small number statistics, is contrary to our expectations based on the result of \citet{Con2008b}.    

%% Why we didn't find closer binaries
   Although the angular resolution of our observations were high enough to resolve binary companions as close as 15~AU in the nearest star forming regions, the two closest companions we found have projected separations of 36~AU (FS Tau A) and 42~AU (IRAS 21004+7811 B).  There are several reasons why we did not find binary companions as close as our detection limit.  Class I YSOs in the closest star forming regions (Taurus, Auriga, Perseus, $\rho$ Ophiuchus) have a lower binary fraction than Class I YSOs in general \citep{Con2008b} and we only observed 5 targets in these regions.  Notably, these two closest binaries are a T Tauri star (FS Tau A) and a star with near-IR colors consistent with a T Tauri star (IRAS 21004+7811, Persi et al. 1988), respectively.  Among our bona fide Class I YSOs, the closest resolved companion has a projected separation of 65~AU from a source in Orion.  Since this is near our resolution limit for targets at the distance of Orion, where many of our targets are located, it is not surprising we did not find closer companions. 

\section{Conclusion}
   We observed 47 Class I YSOs using the NGS-AO system on the Subaru telescope and the LGS-AO system on the Keck II telescope to search for close binary companions to Class I protostars with the goal of testing the hypothesis that protostars with a companion closer than 200~AU should be more likely to have a widely separated companion than protostars without a close companion.  Our observations found 13 companions with a projected separation less than 200~AU.  Contrary to our hypothesis, we found that Class I protostars with a close companion are equally likely to have, or not to have, a widely separated companion within a projected separation of 5,000~AU.  We also found that protostars with such a widely separated companion are not significantly more likely to have a close companion.  However, we found that for \emph{every} protostar with a close companion, there is another YSO within a projected separation of 25,000~AU.  We believe that these protostellar binaries and widely separated YSOs are the remnants of earlier higher order multiple systems that ejected the nearby YSO, now observed at separations up to 25,000~AU.  As such, the observations support our hypothesis.  A third of the close protostellar binaries have a widely separated (and presumably ejected) companion within a projected separation of 5,000~AU that may be gravitationally bound, while the former companions of the other two-thirds now have a projected separation greater than 5,000~AU and are likely to be unbound.  The targets in the Orion star forming region have an excess of close binary companions and an excess of targets with another YSO within 25,000~AU.  This is consistent with Orion having a greater fraction of close binaries than other star forming regions, and the finding that Class I YSOs with a close companion all have another YSO within 25,000~AU.

\acknowledgments
  
  This research has made use of the SIMBAD database, operated at CDS, Strasbourg, France, and NASA's Astrophysics Data System.  This publication makes use of data products from the Two Micron All Sky Survey, which is a joint project of the University of Massachusetts and the Infrared Processing and Analysis Center/California Institute of Technology, funded by the National Aeronautics and Space Administration and the National Science Foundation.  The authors wish to thank Dr. Joshua Walawender and Dr. Sean Matt for providing helpful commentary during the writing of this paper.  The authors wish to recognize and acknowledge the very significant cultural role and reverence that the summit of Mauna Kea has always had within the Hawaiian community.  We are most fortunate to have the opportunity to conduct observations from this mountain. 

\appendix

\section{Comments on Selected Targets}

\textbf{IRAS F04189+2650}:  The close companion to FS Tau A was discovered by \citet{Che1990}. Lunar occultation measurements by \citet{Sim1992} found a separation of 265~mas~$\pm$5~mas and a PA=60$^{\circ}\pm5^{\circ}$.  \citet{Kri1998} used WFPC/HST to measure a separation of 238.7~mas~$\pm$5~mas and a PA=84.4$^{\circ}\pm1.5^{\circ}$. This binary was also observed by \citet{Har2003} with STIS/HST, who measured a separation of 242~mas~$\pm$14~mas and a PA=93.7$^{\circ}\pm2.5^{\circ}$.  The motion of FS Tau Ab relative to FS Tau Aa is shown in Figure 3, which includes our observation.  The path of FS Tau Ab is remarkably linear. Since the proper motion of FS Tau A is to the SSE (PA=167.5) \citep{Duc2005}, the motion of FS Tau Ab should be to the NNW if it were a background star.  However, its motion is too slow and in the wrong direction for it to be a background star given the proper motion of FS Tau A.  Also, spectra taken by \citet{Har2003} show that FS Tau Ab has a late type spectrum with H$\alpha$ emission and thus is unlikely to be a background star.  FS Tau B, 19\farcs75 to the west, is embedded in a reflection nebula and has H-K and K-L$'$ colors consistent with a Class I YSO.
  
\textbf{IRAS 05384$-$0808}:  Four stars in this group are binaries.  The binary star to the west (\#6) is optically visible and thus is not a deeply embedded YSO.  However, it is in a reflection nebula, and so appears to be associated with the embedded stars.  

\textbf{IRAS 06382+1017}: This is a hierarchical triple star.  There is another highly embedded source 13\farcs5 to the east that shares a common nebula with the triple star.  A bright star 17\farcs5 to the west is an unrelated foreground star based on its near-IR colors.  

\textbf{IRAS 07025-1204}: The southern group appears to be a hierarchical triple. 

\textbf{IRAS 18383+0059}:  This is a hierarchical triple where the companion is itself a close binary.  However, the brighter star of the close binary (star Ba in Figure 1) is redder and only slightly fainter than the primary at L$'$, so it may actually be more luminous.  Star Bb is found to be much redder (K-L$'$=3.49) than star Ba (K-L$'$=1.77).

\textbf{IRAS 21004+7811}:  \citet{Per1988} classified this source as a T Tauri star based on J through L photometry.  This is a hierarchical triple where the companion star is itself a 0\farcs14 binary with nearly equal brightness at L$'$.

\textbf{IRAS 21007+4951}:  This deeply embedded (H-K$>$4, K-L$'$=5.06) object is found to have a very close (0\farcs17) companion that is much redder (K-L$'$=7.78) than the primary star.  

\textbf{IRAS 21445+5712}:  We found a possible companion (object B in Figure 1) 0\farcs69 away that is bluer than the primary star (K-L$'$=1.50 vs. K-L$'$=2.56), which is not unusual for companion stars.  At K-band it is small (FWHM=0\farcs09) but elongated, so it is unlikely to be a background star.  It is possible, though unlikely, that this is a very small clump of nebulosity and not a true companion since it is separated from the reflection nebula.

\textbf{IRAS 22376+7455}:  This nearby (300~pc, Hilton \& Lahulla 1995) group has four binary stars.  The close binary to the NW (\#4 in Connelley et al. 2008a) is optically visible and was used as the AO guide star.  Although it is not a deeply embedded YSO, it is in an optical reflection nebula associated with the dark cloud, and thus it is likely to be associated with the embedded stars.

\begin{figure}
\includegraphics[scale=0.90]{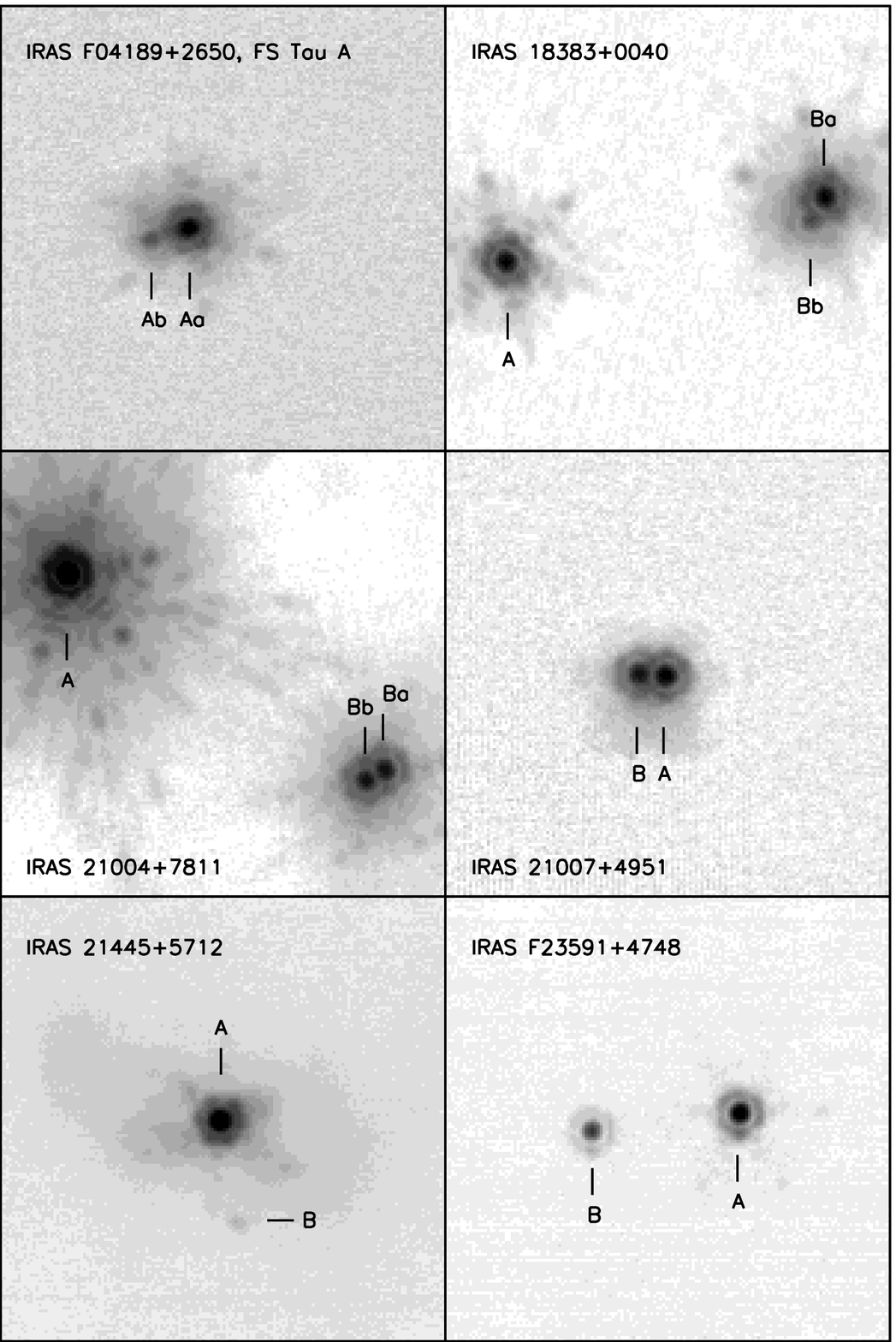}
\caption{3\farcs0$\times$3\farcs0~field of view around each binary observed using the Keck laser guide star AO system at L$'$.  }
\end{figure}

\begin{figure}
\plotone{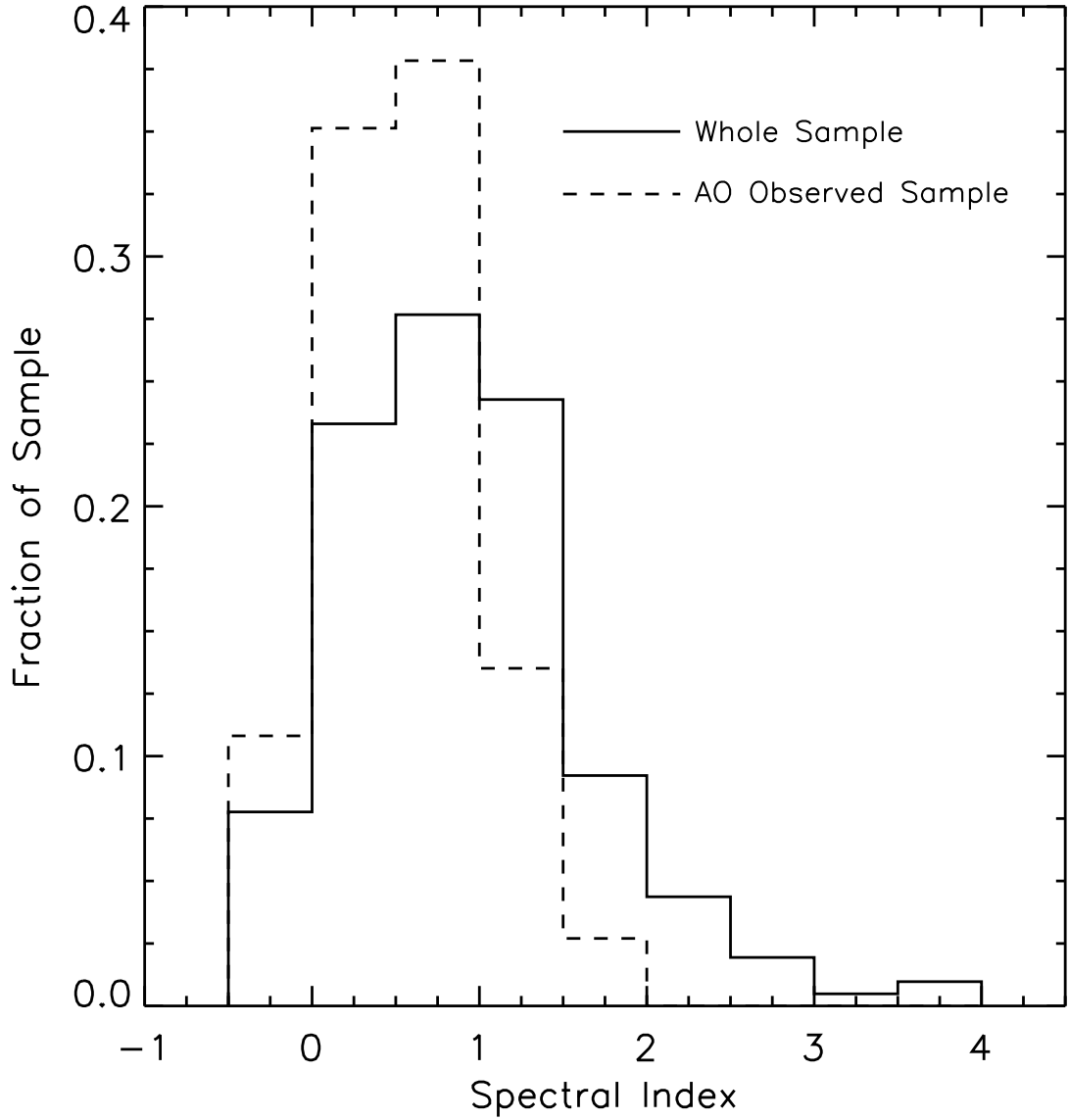}
\caption{Spectral index distribution for the whole sample studied by \citet{Con2008a} versus the same distribution for targets in this study.  The targets in this study are on average younger, but not by a statistically significant margin.  The 2 Sample K-S test shows that these two distributions are consistent with each other within 3~$\sigma$ confidence limits.}
\end{figure}

%% \begin{figure}
%% \plotone{}
%% \caption{}
%% \end{figure}

\begin{figure}
\plotone{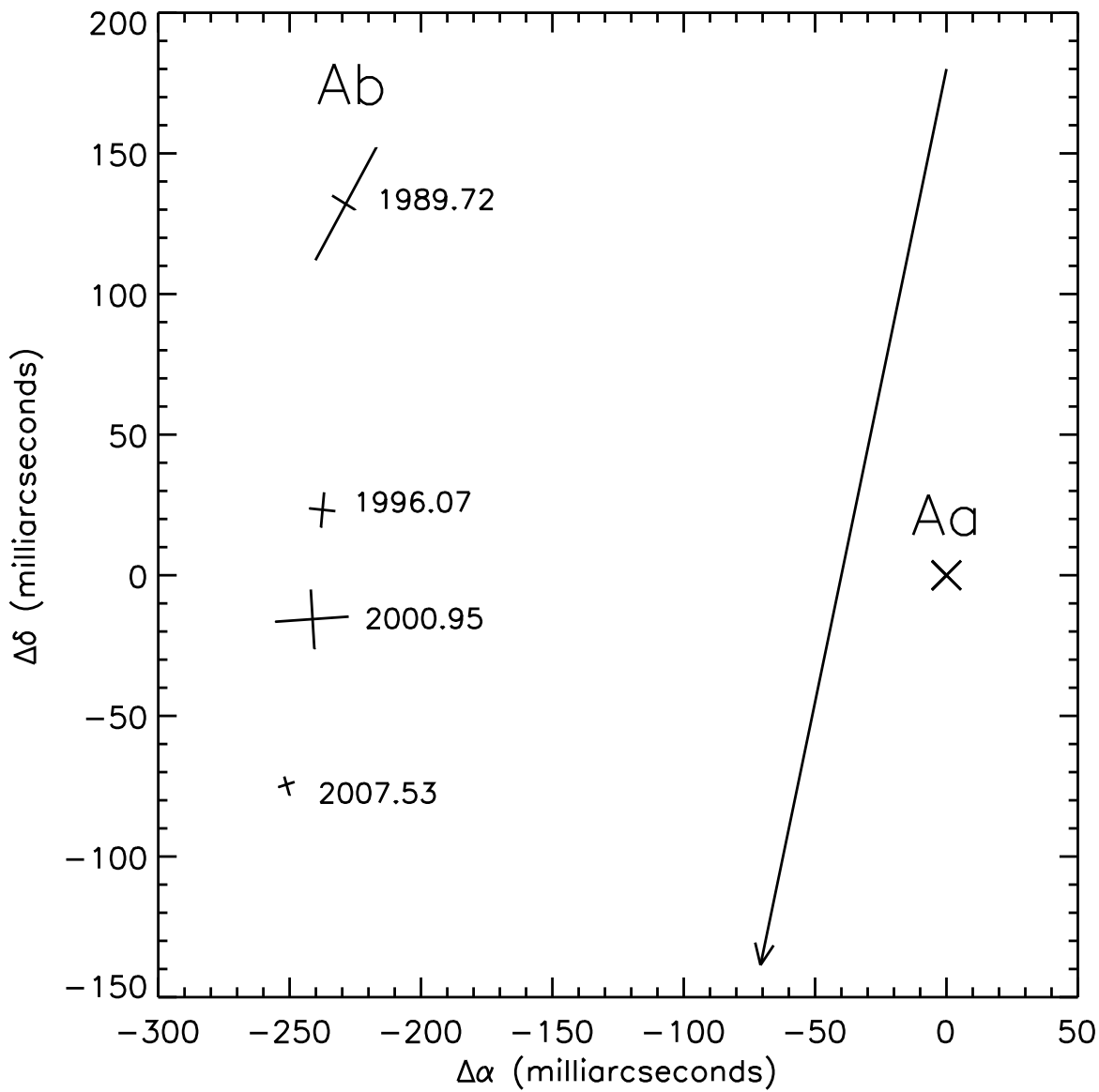}
\caption{The motion of FS Tau Ab relative to FS Tau Aa.  The uncertainties in our position are based on an estimated 1/4 pixel centroiding uncertainty in our image and assume that the image rotator accurately put north up.  The approximate date of the observation is next to each data point.  The spacing of the three most recent measurements is proportional with the time interval between these measurements.  However, the measurement by \citet{Sim1992} in 1989 is too far away, which suggests an error in position angle. The large arrow to the right is the proper motion vector of FS Tau A on the sky in the 17.8 year span of the observations.  FS Tau Aa is located at (0,0) on this figure.}
\end{figure}

%% \begin{figure}
%% \plottwo{}{}
%% \caption{}
%% \end{figure}

\input{AO.table.tex}

\clearpage
\input{AO.table2_new.tex}

\end{document}

%% file: AO.table.tex
%% AO Observed table

\begin{deluxetable}{lccccccccccccc}
\tabletypesize{\scriptsize}
\tablecaption{AO Observed Sample}
\tablewidth{0pt}
\tablecolumns{10}
\tablehead{
\colhead{IRAS} &
\colhead{$\#$\tablenotemark{a}} & \colhead{$\alpha$\tablenotemark{b}(J2000.0)} & \colhead{$\delta$\tablenotemark{b}(J2000.0)} & \colhead{dist~(pc)\tablenotemark{c}} & \colhead{$\Delta$K\tablenotemark{d}} & \colhead{$\Delta$L$'$\tablenotemark{d}} & \colhead{r (\arcsec)} & 
\colhead{PA} & \colhead{UT Date}
}
\startdata

%%  IRAS           num         RA               dec         dist        dK        dL              r               PA             date
03301+3111     & \nodata & 03 33 12.84  &   +31 21 24.1  & 290(2)  & \nodata & \nodata & \nodata         & \nodata         &  09Jul07 \\
F04189+2650    & 2       & 04 22 02.18  &   +26 57 30.5  & 140(3)  & \nodata & 2.67    & 0.264$\pm$0.003 & 106.61$\pm$0.76 &  09Jul07 \\  %% reverified PA  Aug 27, 2007
04240+2559     & \nodata & 04 27 04.70  &   +26 06 16.3  & 140(1)  & \nodata & \nodata & \nodata         & \nodata         &  14Nov05 \\
04530+5126     & \nodata & 04 56 57.02  &   +51 30 50.9  & none    & \nodata & \nodata & \nodata         & \nodata         &  14Nov05 \\
05289$-$0430E  & \nodata & 05 31 27.09  & $-$04 27 59.4  & 417(6)  & \nodata & 2.64    & 0.288$\pm$0.007 & 344.62$\pm$1.48 &  15Nov05 \\  %% verified PA  Aug 7, 2007
05289$-$0430W  & \nodata & 05 31 25.98  & $-$04 28 02.6  & 417(6)  & \nodata & \nodata & \nodata         & \nodata         &  15Nov05 \\
05302$-$0537   & \nodata & 05 32 41.65  & $-$05 35 46.1  & 417(6)  & \nodata & 0.16    & 0.673$\pm$0.006 &  26.44$\pm$0.63 &  14Nov05 \\  %% verified PA  Aug 7, 2007
05327$-$0457W  & 1       & 05 35 13.10  & $-$04 55 52.5  & 417(6)  & \nodata & 1.23    & 0.147$\pm$0.007 &  82.18$\pm$2.90 &  15Nov05 \\  %% verified PA  Aug 7, 2007
05357$-$0650   & \nodata & 05 38 09.31  & $-$06 49 16.6  & 417(6)  & \nodata & \nodata & \nodata         &  \nodata        &  15Nov05 \\
05375$-$0040   & 1       & 05 40 06.79  & $-$00 38 38.1  & 417(6)  & \nodata & 1.68    & 6.403$\pm$0.007 & 278.97$\pm$0.07 &  14Nov05 \\  %% verified PA  Aug 7, 2007
05384$-$0808   & 1       & 05 40 50.59  & $-$08 05 48.7  & 417(6)  & \nodata & 1.81    & 0.377$\pm$0.005 & 141.51$\pm$1.13 &  15Nov05 \\  %% verified PA  Aug 7, 2007
05384$-$0808   & 3       & 05 40 49.92  & $-$08 06 08.4  & 417(6)  & \nodata & 1.89    & 0.178$\pm$0.007 & 355.25$\pm$2.40 &  15Nov05 \\  %% verified PA  Aug 7, 2007
%%05384$-$0807 & 4       & 05 40 50.05  & $-$08 05 55.2   & 480(1) & \nodata  & 0.78    & 0.08 & 327.5 &  15Nov05 \\  %% verified PA  Aug 7, 2007
05384$-$0808   & 6       & 05 40 48.07  & $-$08 05 58.7  & 417(6)  & \nodata & 0.73    & 0.162$\pm$0.005 & 314.09$\pm$2.64 &  15Nov05 \\  %% verified PA  Aug 7, 2007
05404$-$0948   & 1       & 05 42 47.70  & $-$09 47 22.2  & 417(6)  &  3.48   & 2.49    & 3.594$\pm$0.006 & 204.92$\pm$0.12 &  15Nov05 \\  %% verified PA  Aug 7, 2007
05404$-$0948   & 1       & 05 42 47.70  & $-$09 47 22.2  & 417(6)  &  2.57   & 2.36    & 0.162$\pm$0.005 & 135.32$\pm$2.63 &  15Nov05 \\  %% verified PA  Aug 7, 2007
%%  IRAS           num         RA               dec         dist        dK        dL              r               PA             date
05513$-$1024   & \nodata & 05 53 42.55  & $-$10 24 00.7  & 417(6)  & \nodata & \nodata & \nodata         & \nodata         &  15Nov05 \\
05555$-$1405N  & 1       & 05 57 49.46  & $-$14 05 27.8  & 417(6)  & \nodata & 0.08    & 5.797$\pm$0.007 & 177.52$\pm$0.07 &  15Nov05 \\  %% verified PA  Aug 7, 2007
05555$-$1405N  & 1       & 05 57 49.46  & $-$14 05 27.8  & 417(6)  & \nodata & 1.55    & 0.209$\pm$0.006 & 115.31$\pm$2.04 &  15Nov05 \\  %% verified PA  Aug 7, 2007
05555$-$1405S  & 4       & 05 57 49.18  & $-$14 06 08.0  & 417(6)  & \nodata & \nodata & \nodata         & \nodata         &  15Nov05 \\
06297+1021W    & \nodata & 06 32 26.12  &   +10 19 18.4  & 900(2)  & \nodata & \nodata & \nodata         & \nodata         &  15Nov05 \\
06382+1017     & 1       & 06 41 02.64  &   +10 15 02.4  & 800(3)  & \nodata & 2.46    & 1.818$\pm$0.006 & 336.56$\pm$0.23 &  14Nov05 \\  %% verified PA  Aug 7, 2007
06382+1017     & 1       & 06 41 02.64  &   +10 15 02.4  & 800(3)  & \nodata & 1.85    & 0.169$\pm$0.007 &  17.20$\pm$3.12 &  14Nov05 \\  %% verified PA  Aug 7, 2007
07025$-$1204   & 2       & 07 04 51.62  & $-$12 09 29.9  & 1150(7) & \nodata & 0.43    & 0.345$\pm$0.006 & 330.32$\pm$1.24 &  15nov05 \\  %% verified PA  Aug 7, 2007
07025$-$1204   & 2       & 07 04 51.62  & $-$12 09 29.9  & 1150(7) & \nodata & 1.56    & 2.364$\pm$0.005 & 142.76$\pm$0.18 &  15nov05 \\  %% verified PA  Aug 7, 2007
07025$-$1204   & 4       & 07 04 50.13  & $-$12 09 15.7  & 1150(7) & \nodata & 3.58    & 1.488$\pm$0.007 & 356.69$\pm$0.29 &  15nov05 \\  %% verified PA  Aug 7, 2007
07025$-$1204   & 4       & 07 04 50.13  & $-$12 09 15.7  & 1150(7) & \nodata & 4.45    & 0.638$\pm$0.007 & 355.13$\pm$0.67 &  15nov05 \\  %% verified PA  Aug 7, 2007
08043$-$3343   & \nodata & 08 06 15.61  & $-$33 52 19.5  & 1120(4) & \nodata & \nodata & \nodata         & \nodata         &  15Nov05 \\
16235$-$2416   & \nodata & 16 26 34.17  & $-$24 23 28.3  & 160(1)  & \nodata & \nodata & \nodata         & \nodata         &  10Jun07 \\
16316$-$1540   & \nodata & 16 34 29.31  & $-$15 47 01.4  & 160(1)  & \nodata & \nodata & \nodata         & \nodata         &  10Jun07 \\
18250$-$0351   & \nodata & 18 27 39.53  & $-$03 49 52.0  & 280(8)  & \nodata & \nodata & \nodata         & \nodata         &  10Jun07 \\
18275+0040     & \nodata & 18 30 06.17  &   +00 42 33.6  & 700(9)  & \nodata & \nodata & \nodata         & \nodata         &  10Jun07 \\
18383+0059     & 3       & 18 40 51.73  &   +01 02 13.2  & none    &  0.81   & 0.05    & 2.194$\pm$0.003 & 280.61$\pm$0.56 &  10Jun07 \\  %% verified PA  Aug 7, 2007. PA is from UKIRT/UIST data.  PA as measured by Keck data is 296.43
18383+0059     & 1       & 18 40 51.88  &   +01 02 12.8  & none    &  4.19   & 2.47    & 0.184$\pm$0.003 & 147.70$\pm$1.23 &  10Jun07 \\  %% reverified PA  Aug 27, 2007, measured PA from Keck data, then applied difference in PA measured for wide binary between UKIRT and Keck data
19247+2238     & 1       & 19 26 51.33  &   +22 45 13.4  & none    & \nodata & 0.69    & 1.569$\pm$0.006 & 143.70$\pm$0.27 &  14Nov05 \\  %% verified PA  Aug 7, 2007
19247+2238     & 3       & 19 26 51.62  &   +22 45 03.2  & none    & \nodata & 0.97    & 0.260$\pm$0.007 &  18.70$\pm$1.64 &  14Nov05 \\  %% verified PA  Aug 7, 2007
19266+0932     & \nodata & 19 29 00.86  &   +09 38 42.9  & 300(3)  & \nodata & \nodata & \nodata         & \nodata         &  09Jul07 \\
%%  IRAS           num         RA               dec         dist        dK        dL              r               PA             date
20355+6343     & \nodata & 20 36 22.86  &   +63 53 40.4  & 450(5)  & \nodata & \nodata & \nodata         & \nodata         &  10Jun07 \\
20453+6746     & \nodata & 20 38 57.48  &   +57 09 37.6  & 500(3)  & \nodata & \nodata & \nodata         & \nodata         &  14Nov05 \\
20568+5217     & \nodata & 20 58 21.09  &   +52 29 27.7  & 1270(4) & \nodata & \nodata & \nodata         & \nodata         &  15Nov05 \\
21004+7811     & 1       & 20 59 03.73  &   +78 23 08.8  & 300(3)  &  2.50   & 3.12    & 2.515$\pm$0.002 & 235.35$\pm$0.47 &  09Jul07 \\ %% reverified PA  Aug 27, 2007  Contrasts are between primary and brighter member of binary.  Listed PA was measured from IRCS data.  PA measured from Keck data is 225.276
21004+7811     & 2       & 20 59 03.13  &   +78 23 08.6  & 300(3)  &  0.35   & 0.01    & 0.141$\pm$0.003 & 119.34$\pm$1.50 &  09Jul07 \\ %% PA as measured from Keck data is 109.268.  Applied difference bewteen IRCS and Keck images
21007+4951     & \nodata & 21 02 22.67  &   +50 03 08.1  & 700(1)  &  3.45   & 0.78    & 0.176$\pm$0.003 &  86.75$\pm$1.14 &  09Jul07 \\ %% verified PA   Aug 27, 2007
21388+5622     & 2       & 21 40 28.02  &   +56 36 02.8  & 750(10) & \nodata & 0.63    & 0.711$\pm$0.005 & 133.36$\pm$0.60 &  15Nov05 \\ %% reverified PA  Aug 27, 2007
21388+5622     & 3       & 21 40 28.06  &   +56 36 06.1  & 750(10) & \nodata & 1.34    & 2.495$\pm$0.005 &  47.54$\pm$0.17 &  15Nov05 \\ %% verified PA  Aug 7, 2007
21445+5712     & \nodata & 21 46 07.12  &   +57 26 31.8  & 360(4)  &  4.11   & 5.17    & 0.694$\pm$0.003 & 189.68$\pm$0.29 &  10Jun07 \\ %% verified PA  Aug 27, 2007, some confusion over image orientation of Keck images
21454+4718     & \nodata & 21 47 20.66  &   +47 32 03.6  & 900(1)  & \nodata & \nodata & \nodata         & \nodata         &  15Nov05 \\
21454+4718     & \nodata & 21 47 20.66  &   +47 32 03.6  & 900(1)  & \nodata & \nodata & \nodata         & \nodata         &  09Jul07 \\
21569+5842     & \nodata & 21 58 35.90  &   +58 57 22.8  & 250(4)  & \nodata & \nodata & \nodata         & \nodata         &  09Jul07 \\
F22324+4024    & \nodata & 22 34 41.01  &   +40 40 04.5  & 880(3)  & \nodata & \nodata & \nodata         & \nodata         &  09Jul07 \\
22376+7455     & 1       & 22 38 47.02  &   +75 11 34.7  & 330(1)  & \nodata & 2.22    & 0.545$\pm$0.007 & 175.94$\pm$0.78 &  15Nov05 \\ %% verified PA  Aug 7, 2007
22376+7455     & 2       & 22 38 43.98  &   +75 11 26.9  & 330(1)  & \nodata & 0.49    & 0.485$\pm$0.007 & 340.69$\pm$0.88 &  15Nov05 \\ %% verified PA  Aug 7, 2007
22376+7455     & 4       & 22 38 42.49  &   +75 11 45.6  & 330(1)  & \nodata & 0.95    & 0.695$\pm$0.007 & 266.78$\pm$0.61 &  15Nov05 \\ %% verified PA  Aug 22, 2007
22376+7455     & 6       & 22 38 42.83  &   +75 11 37.0  & 330(1)  & \nodata & 0.16    & 1.232$\pm$0.005 & 142.27$\pm$0.35 &  15Nov05 \\
F23591+4748    & \nodata & 00 01 43.25  &   +48 05 19.0  &  none   &  1.64   & 1.57    & 1.004$\pm$0.003 &  96.81$\pm$0.20 &  10Jun07 \\ %% reverified PA  Aug 27, 2007. PA measured in Keck data agrees w/ PA from UKIRT data in paper #1

\enddata
\tablenotetext{a}{The number of the primary star in the finder charts presented in \citet{Con2008a}.}
\tablenotetext{b}{Coordinates are from 2MASS, or offset from a field star with 2MASS coordinates}
\tablenotetext{c}{The estimated distance to each source in parsecs.  The citation for the distances estimate is designated by the number in the parentheses, and are as follows: 1) Hilton, J., \& Lahulla, J. 1995; 2) Educated guess based on proximity to nearby objects; 3) Reipurth, A General Catalog of HH Objects, 1999 (http:casa.colorado.edu/hhcat/); 4) Wouterloot \& Brand, 1989; 5) Launhardt \& Henning 1997; 6) Menten et al. 2007; 7) Sugitani \& Ogura 1995; 8) Bachiller et al. 2001; 9) Zhang et al. 1988; 10) Battinelli \& Capuzzo-Dolcetta 1991}
\tablenotetext{d}{The observed magnitude difference between the two binary components in the specified band.}

\end{deluxetable}

%% file: AO.table2_new.tex
%% AO Results Table
%% Note: Aug 20, 2007: Verify all numbers after I hear back from Bo about including optically visible binaries
%%                     Make sure list of targets used for these stats and list of binaries in table 1 match, or else say why they don't
\rotate{
\begin{deluxetable}{lccccccccccccc}
\tabletypesize{\scriptsize}
\tablecaption{AO Survey Results, Resticted Sample}
\tablewidth{0pt}
\tablecolumns{6}
\tablehead{
\colhead{} & \multicolumn{2}{c}{X=5,000}  & \multicolumn{2}{c}{X=25,000} \\
\colhead{Question} &
\colhead{$\#$} & \colhead{$\%$\tablenotemark{a}} & \colhead{$\#$} & \colhead{$\%$\tablenotemark{a}}
}
\startdata

%%  Question      numbers     fraction   numbers    fraction    answer
1) Are protostars with a wide companion within X~AU more likely to have a close companion            &      &                            &      &                             \\
than a protostar without a wide companion within X~AU?                                               &      &                            &      &                             \\
~~Protostars with a wide companion and have a close companion:                                       & 3/10 & $30.0\%^{+20.7\%}_{-15.8\%}$  & 9/19 & $47.3\%^{+13.8\%}_{-13.4\%}$  \\
~~Protostars without a wide companion that have a close companion:                                   & 6/22 & $27.3\%^{+12.7\%}_{-10.2\%}$  & 0/13 & $0.0\%^{+13.2\%}_{-0.0\%}$    \\
2) Are protostars with a close companion more likely to have a wide                                  &      &                            &      &                             \\
companion within X AU?                                                                               &      &                            &      &                             \\
~~Protostars with a close companion but without widely separated YSO within X~AU                     & 6/9  & $66.7\%^{+17.5\%}_{-22.0\%}$  & 0/9  & $0.0\%^{+18.4\%}_{-0.0\%}$    \\
~~Protostars with a close companion and with a widely separated YSO within X~AU                      & 3/9  & $33.3\%^{+22.0\%}_{-17.5\%}$  & 9/9  & $100.0\%^{+0.0\%}_{-18.4\%}$  \\
3) Are protostars without a close companion less likely to have an apparently widely                 &      &                            &      &                             \\
separated YSO within X~AU?                                                                           &      &                            &      &                             \\
~~Protostars without a close companion and without a widely separated YSO within X~AU?               & 16/23& $69.6\%^{+10.4\%}_{-12.4\%}$  & 13/23&  $56.5\%^{+11.8\%}_{-12.5\%}$ \\
~~Protostars without a close companion but with a widely separated YSO  within X~AU?                 & 7/23 & $30.4\%^{+12.4\%}_{-10.4\%}$  & 10/23&  $43.5\%^{+12.5\%}_{-11.8\%}$ \\
4) Does the fraction of targets with a wide companion have a regional dependence?                    &      &                            &      &                             \\
~~Number with a widely separated YSO within X~AU in the Orion sub-sample                             & 4/12 & $33.3\%^{+18.5\%}_{-15.1\%}$  & 10/12&  $83.3\%^{+10.7\%}_{-17.8\%}$ \\
~~Number with a widely separated YSO within X~AU in the non-Orion sub-sample                         & 6/20 & $30.0\%^{+13.5\%}_{-11.1\%}$  & 10/20 &  $50.0\%^{+13.2\%}_{-13.2\%}$ \\
\enddata

\tablenotetext{a}{1~$\sigma$ uncertainties are presented, derived using binomial statistics with the calculator at http://statpages.org/confint.html}

\end{deluxetable}
}    %% end of \rotate command